\documentclass[twocolumn,showpacs,superscriptaddress,floatfix,prb]{revtex4}
\usepackage{graphicx,amsfonts,amssymb,amsmath,hyperref,hypcap,enumerate}

\begin{document}
\title{Exciton band structure of monolayer MoS$_2$}

\author{Fengcheng Wu}
\affiliation{Department of Physics, University of Texas at Austin, Austin, TX 78712, USA}

\author{Fanyao Qu}
\email{fanyao@unb.br}
\affiliation{Department of Physics, University of Texas at Austin, Austin, TX 78712, USA}
\affiliation{Instituto de F\'isica, Universidade de Bras\'ilia, Bras\'ilia-DF 70919-970, Brasil}

\author{A. H. MacDonald}
\email{macdpc@physics.utexas.edu}
\affiliation{Department of Physics, University of Texas at Austin, Austin, TX 78712, USA}

\date{\today}

\begin{abstract}
We address the properties of excitons in monolayer MoS$_2$ from a theoretical point of view,
showing that low-energy excitonic states
occur both at the Brillouin zone center and at the Brillouin-zone corners, that
binding energies at the Brillouin-zone center deviate strongly
from the $(n-1/2)^{-2}$ pattern of the two-dimensional hydrogenic
model, and that the valley-degenerate exciton doublet at the Brillouin-zone center
splits at finite momentum into an upper mode with non-analytic linear dispersion and a lower
mode with quadratic dispersion.  Although monolayer MoS$_2$ is a direct-gap semiconductor when
classified by its quasiparticle band structure, it may well be an indirect gap material when classified by its
excitation spectra.
\end{abstract}

% 71.35.-y Excitons and related phenomena
% 78.67.-n Optical properties of low-dimensional, mesoscopic, and nanoscale materials and structures
% 73.22.-f Electronic structure of nanoscale materials and related systems

\maketitle

\section{Introduction}

In monolayer form the group VI transition-metal dichalcogenides (TMDs) like MoS$_2$
are an interesting class of semiconductors, and one
that has recently received considerable attention.
\cite{Wang12, Heinz14, Yao12, VP_Peking, VP_HK, VP_Columbia,PL10_Berkeley, PL10_Columbia, PL_Shih, PL_Rana, Trion_Columbia, Trion_Seattle, Trion_Austin, GW12, Steven13, VHall, Tutuc14}
In these materials conduction and valence bands are both dominantly $d$-electron in character and
have band extrema located at the triangular lattice Brillouin-zone-corners $K$ and $K'$.
Because their structure breaks inversion symmetry, coupling is allowed between real spin and
valley pseudospin\cite{Yao12} and gives rise to valley-dependent optical selection
rules.\cite{VP_Peking, VP_HK, VP_Columbia}
Because of relatively large carrier effective masses, reduced screening, and carrier confinement in a single atomic layer,
their electron-hole interactions are much stronger than in conventional semiconductors.
Monolayer TMDs therefore host exceptionally strongly bound excitons and
trions that have been extensively studied both experimentally and theoretically.\cite{PL10_Berkeley, PL10_Columbia, PL_Shih, PL_Rana, Trion_Columbia, Trion_Seattle, Trion_Austin, GW12, Steven13}

In this article we report on a theoretical study of exciton energies and wave functions in MoS$_2$
as a function of momentum across the full Brillouin zone.
We identify important aspects of 2D-TMD exciton
physics that are controlled by mirror, three-fold rotational, and time-reversal discrete symmetries.
We calculate the optical conductivity, which reflects the properties of excitons with zero center-of-mass momentum
and exhibits a set of peaks split by electron-hole binding energies as usual, but also by
large valence band spin-orbit coupling energies. The spin splitting of valence band
is conventionally used to classify absorption peaks into $A$ and $B$ series.
The exciton energy pattern is distinctly different from that of a 2D hydrogenic model.
In particular the four $A_{2p}$ states are lower in energy than the corresponding $2s$ states,
and not degenerate.

Finite-momentum excitons are optically inactive, but can nevertheless play an important role in
hot carrier relaxation and in valley dynamics.\cite{Dynamics13, Mai13, Mai14, Urbaszek14, Wu14, Marie14}
We find that for both $A$ and $B$ excitons, the valley degenerate states at the Brilliouin-zone
center split at small momentum into a lower mode with quadratic dispersion and an upper mode
with non-analytic linear dispersion.  This unusual pattern is due to valley coherence
established by electron-hole exchange interactions.

Low energy exciton states appear both near the Brillouin-zone center and near the Brillouin-zone corners.
There are two distinct types of Brillouin-zone corner excitons.
One type has electrons and holes in opposite valleys ($K$, $K'$), while the other has holes in the  $\Gamma$ valley and
electrons in either the $K$ or $K'$ valley.  Although monolayer MoS$_2$ is a direct-gap semiconductor
as judged by its band structure, because of these Brillouin-zone corner exciton states,
we propose that it may well be an indirect gap material when judged by its excitation spectra.

Our paper is organized as follows.  In Section II we describe the model we employ for quasiparticle bands
and for electron-hole interactions, and in Section III we present our results.  We conclude in Section IV with a
summary and brief discussion.

\begin{figure}[t]
 \includegraphics[width=1\columnwidth]{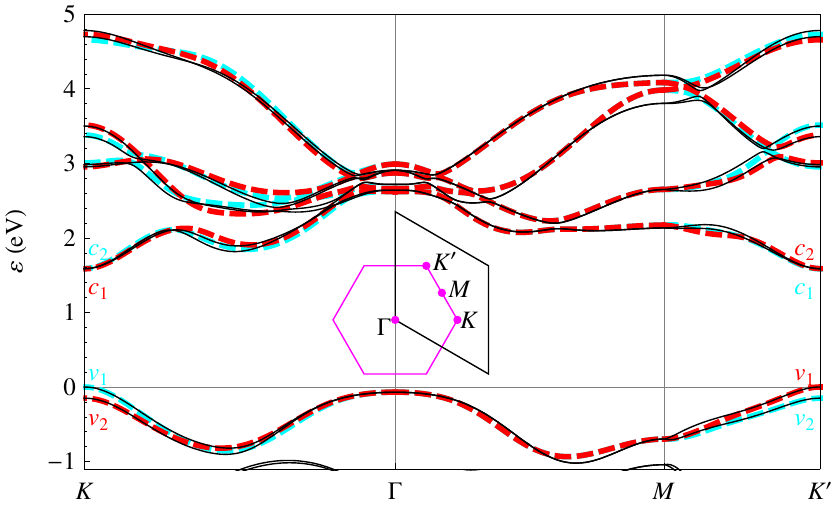}
 \caption{(Color online) Quasiparticle band structure of monolayer MoS$_2$.
  The solid curves were obtained using the \textsc{Quantum ESPRESSO} package\cite{qespresso}
  with fully relativistic pseudopotentials under the Perdew-Burke-Ernzerhof generalized-gradient approximation, and a $16\times16\times1$ $\vec{k}$-grid.
  The dashed curves were calculated from the tight-binding model,
  with cyan (red) representing states that are even (odd) under mirror operation with respect to the Mo plane.
  $v_{1,2}$ and $c_{1,2}$ label the bands close to the valence and conduction band edges
  near the $K$ and $K'$ points.  The inset shows the hexagonal
  Brillouin zone (pink) associated with the triangular Bravais lattice of MoS$_2$ and an alternate rhombohedral primitive zone (black),
  and labels the principle high-symmetry points in reciprocal space.  Note that the valence band
  maxima at $\Gamma$ is only slightly lower in energy than the valence band maxima at $K,K'$}
 \label{band}
\end{figure}

\begin{figure*}[t]
 \includegraphics[width=2.0\columnwidth]{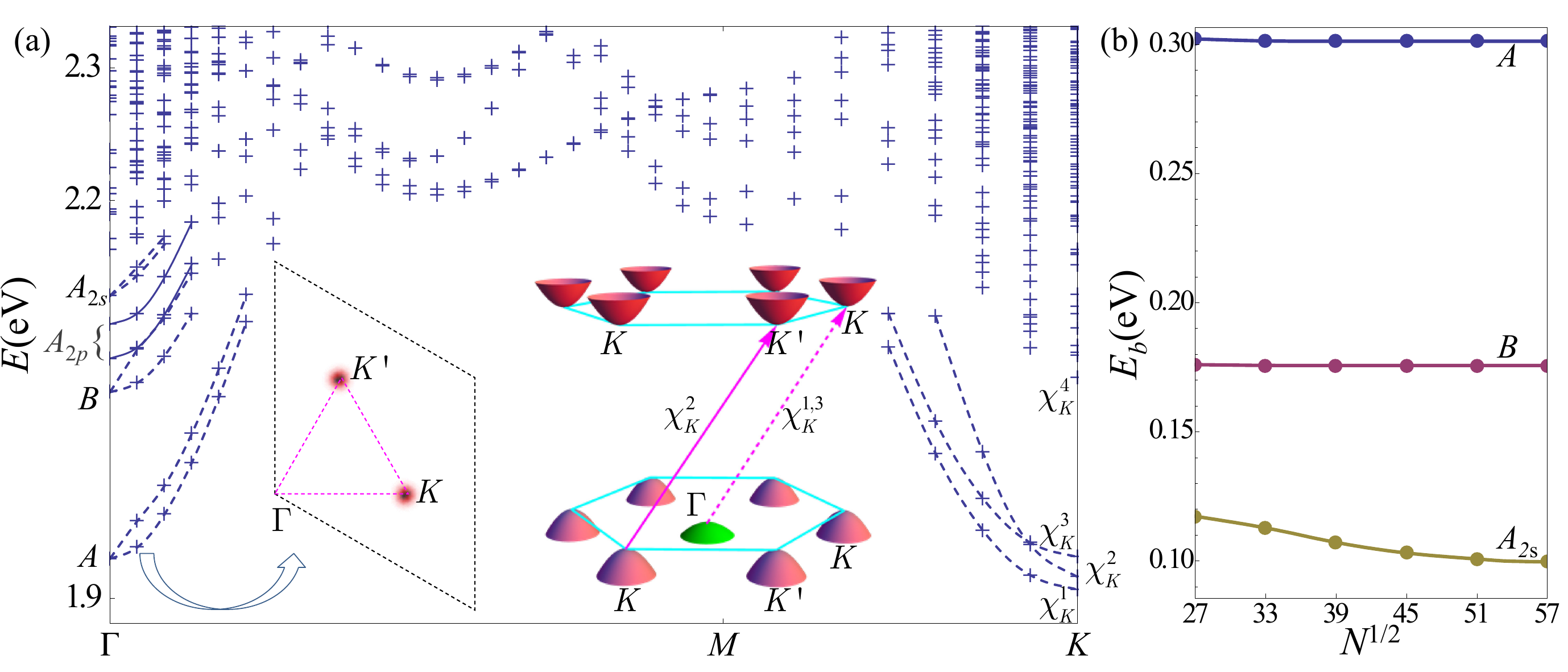}
 \caption{(Color online) (a) Energies of Type III excitons as a function of center-of-mass momentum $\vec{Q}$.
  This figure is based on a calculation performed using a $45\times 45$ $\vec{k}$-grid.
  The lines were added as a guide to the eye. Solid (dashed) lines represent states that are doubly(singly) degenerate.
  The labels of the excitons with $\vec{Q}=0$ are explained in the main text.
  Excitons with $\vec{Q}= K$  are labeled by $\chi_{K}^1$, $\chi_{K}^2$ and so on
  in ascending order of energy.    The left inset is a $\vec{k}$-space map plot of $P_{\vec{Q}}(\vec{k})$
  (see Eq.~(\ref{Prob})) for the $\vec{Q}=\frac{2}{45} M$ exciton
  in the lower-energy branch evolving from $A$.
  The right inset schematically illustrates the dominant
   electron-hole transitions which contribute to the $\chi_{K}^1$, $\chi_{K}^2$ and $\chi_{K}^3$ exciton states.
  (b) Binding energy $E_b$ for $A$, $B$ and $A_{2s}$ excitons at $\vec{Q}=0$
   as a function of $N^{1/2}$, where $N$ is number of $\vec{k}$ points.}
 \label{Exciton}
\end{figure*}

\section{Theoretical Formulation}

Exciton states are obtained by solving a two-particle problem with
attractive interactions between one conduction band electron and one valence band hole.
Because the valence and conduction band edges are dominated by Mo atom $d$-orbitals
we use a five band $d$-orbital tight-binding model, detailed in Appendix~\ref{AppA} and illustrated in
Fig.~\ref{band}, for the quasiparticle bands
of the TMD semiconductor ground state.  Provided that the typical
separation between the electrons and holes in exciton states is at least several lattice
constants we can assume that the electron-hole interaction strengths
are dependent mainly on the separation between atomic sites and not on the $d$-orbital character
on that site.  These considerations lead to a Hamiltonian of the form
 \begin{equation}
\begin{aligned}
H&=H_{\text{0}}+H_{\text{I}},\\
H_{\text{I}}&=\frac{1}{2}\sum_{\vec{R}, \vec{R}'}V_{|\vec{R}-\vec{R}'|}
a^{\dagger}_{\vec{R}\nu}
a^{\dagger}_{\vec{R}'\nu'}
a_{\vec{R}'\nu'}
a_{\vec{R}\nu}.
\end{aligned}
\label{Ham}
\end{equation}
where $H_{\text{0}}$ is the Hamiltonian for independent $d$ electrons and $H_{\text{I}}$ describes their interactions.
In Eq.~(\ref{Ham}) $a^\dagger_{\vec{R}\nu}$ ($a_{\vec{R}\nu}$) is the electron creation (annihilation)
operator for orbital $\nu$ at Mo site $\vec{R}$ and $\nu=(o,s)$ includes both orbital $o$ and spin $s$ labels.
We follow recent work\cite{Reichman, Heinz_14} by using
an interaction potential of the Keldysh form,\cite{Keldysh, Rubio}
\begin{equation}
V_{R}=\frac{\pi e^2}{2 \epsilon r_0} [H_0(R/r_0)-Y_0(R/r_0)],
\label{potential}
\end{equation}
to account for the finite width of the TMD layer and the spatial inhomogeneity
of the dielectric screening environment.  This interaction gives a good description of the
nonhydrogenic Rydberg series observed in monolayer WS$_2$.\cite{Reichman, Heinz_14}
In Eq.~(\ref{potential}) $\epsilon$ is an environment-dependent dielectric constant,
$r_0$ is a characteristic length related to the width of a single TMD layer, and $H_0$ and $Y_0$ are respectively
Struve and Bessel functions of the second kind.
Unless otherwise stated, we chose $\epsilon=2.5$, which corresponds to MoS$_2$ lying on a SiO$_2$ substrate
and exposed to air. $r_0$ depends on $\epsilon$ and we took $r_0=33.875\text{\AA}/\epsilon$ from Ref.~[\onlinecite{PL_Rana}] .
The onsite interaction is regularized by setting $V_{0}= U V_{R=a_0}$ with $a_0$
equal to the lattice parameter of MoS$_2$, and the parameter $U$ is taken to be 1 for results presented below.
The dependence of our results on the value chosen for the dimensionless parameter $U$,
which accounts for screening of on-site potentials by remote bands, will be discussed later.

Exciton states with center of mass momentum $\vec{Q}$ can be expanded in terms of one-electron/one-hole states:
\begin{equation}
|\chi\rangle_{\vec{Q}}=\sum_{v, c, \vec{k}} \psi_{\vec{Q}}(v,c,\vec{k}) \; |v,c,\vec{k},\vec{Q}\rangle,
\end{equation}
where $|v,c,\vec{k},\vec{Q}\rangle=b^{\dagger}_{(\vec{k}+\vec{Q})c}b_{\vec{k}v}|G\rangle$,
$|G\rangle$ is the neutral semiconductor ground state, and the sums are over all valance ($v$) and conduction ($c$) bands.
$b_{\vec{k} n}$ and $b_{\vec{k} n}^\dagger$ are quasiparticle operators for band $n$ at momentum $\vec{k}$.
The wave vector $\vec{k}+\vec{Q}$ is understood to be reduced to the Brillouin-zone.
The exciton center-of-mass momentum $\vec{Q}$ is also understood to be confined to the
Brillouin-zone and is a good quantum number.  Like the quasiparticles, excitons have a
band structure. To characterize an exciton state, we define its $\vec{k}$-space probability distribution function as
\begin{equation}
P_{\vec{Q}}(\vec{k})=\sum_{v, c}|\psi_{\vec{Q}}(v,c,\vec{k})|^2.
\label{Prob}
\end{equation}

The eigenvalue problem for the Hamiltonian matrix projected onto this subspace
is a Bethe-Salpeter (BS) equation.  Its solution determines the
exciton energies  $E_{\vec{Q}}$ and wave functions.  The Hamiltonian matrix
\begin{equation}
\begin{aligned}
&\langle v,c,\vec{k},\vec{Q}|H|v',c',\vec{k}',\vec{Q}\rangle\\
=&\delta_{vv'}\delta_{cc'}\delta_{\vec{k}\vec{k}'}\big(\varepsilon_{(\vec{k}+\vec{Q})c}-\varepsilon_{\vec{k}v}\big)
-\big(D-X\big)_{vv'}^{cc'}(\vec{k},\vec{k}',\vec{Q}),
\end{aligned}
\label{BS_matrix}
\end{equation}
where $\varepsilon_{\vec{k}n}$ denotes the quasiparticle energy.
We view the two-dimensional bands predicted by density-functional theory electronic structure calculations,
illustrated in Fig.~\ref{band} as solutions of the neutral semiconductor single-particle Dyson equation including
all many-body self-energy effects except for finite lifetimes.  It follows that in the exciton calculation we need to
account only for corrections due to electron-hole interactions.  It will, however, be necessary to correct for the
well-known tendency of density-functional-theory bands to underestimate semiconductor gaps.
As discussed later, this consideration motivates shifting the calculated
excitation energy spectrum rigidly to match experimental optical absorption spectra.

In Eq.~(\ref{BS_matrix}), $D$ and $X$ are respectively the direct and exchange two-particle matrix elements:
\begin{equation}
\begin{aligned}
&D_{vv'}^{cc'}(\vec{k},\vec{k}',\vec{Q})
=\frac{1}{N}V_{\vec{k}-\vec{k}'}
\big(\mathcal{U}^\dagger_{\vec{k}+\vec{Q}}\mathcal{U}_{\vec{k}'+\vec{Q}}\big)_{cc'}\big(\mathcal{U}^\dagger_{\vec{k}'}\mathcal{U}_{\vec{k}}\big)_{v'v},\\
&X_{vv'}^{cc'}(\vec{k},\vec{k}',\vec{Q})
=\frac{1}{N}V_{\vec{Q}}
\big(\mathcal{U}^\dagger_{\vec{k}+\vec{Q}}\mathcal{U}_{\vec{k}}\big)_{cv}\big(\mathcal{U}^\dagger_{\vec{k}'}\mathcal{U}_{\vec{k}'+\vec{Q}}\big)_{v'c'},\\
\end{aligned}
\label{DX}
\end{equation}
where $\mathcal{U}_{\vec{k}}$ is the unitary matrix which diagonalizes the quasiparticle Hamiltonian (see Appendix~\ref{AppA}),
$N$ is the number of unit cells in the finite system over which we apply periodic boundary conditions,
and $V_{\vec{q}}=\sum e^{i\vec{q}\cdot\vec{R}}V_{R}$ is the lattice Fourier transform of the interaction potential.
Note that $V(\vec{q})= V(\vec{q}+\vec{G})$ for any reciprocal lattice vector $\vec{G}$.
For the special case $\vec{Q}=0$, the exchange term $X$ vanishes
because of the orthogonality property $\big(\mathcal{U}^\dagger_{\vec{k}}\mathcal{U}_{\vec{k}}\big)_{cv}=0$.
We remark that exchange term survives even at $\vec{Q}=0$ in models in which electron-hole
interactions depend not only on electron-hole separation, but also on orbital character\cite{Wu14}.

\section{Exciton Band Structure}

Monolayer MoS$_2$ has mirror symmetry \cite{mirror} with respect to the Mo plane.
Quasiparticle band spinors can be classified by this symmetry $\mathcal{M}|\vec{k},n\rangle=-i m_{n}|\vec{k},n\rangle$,
where the mirror number $m_{n}=+(-)$ for mirror even (odd) bands as shown in Fig.~\ref{band}.
Using mirror numbers, we can group exciton states into three decoupled types:
(1) A Type I exciton is formed by promoting an electron from a mirror-even
valance band $(m_v=+)$ to mirror-odd conduction bands $(m_c=-)$;
(2) Type II is similar to Type I but with $(m_v, m_c)=(-, +)$;
(3) For a Type III exciton, $m_v=m_c=\pm$.
Only Type III excitons can be optically bright.
Exchange terms vanish in Type I and II excitons
because their valence and conduction bands have opposite mirror numbers.
For a Type III exciton, the two sectors $m_v=m_c=+$ and $m_v=m_c = -$ are coupled by exchange terms, but not by direct terms.
In the following, we restrict our discussion to Type III excitons, although many of the points we make apply equally well to
Type I and Type II excitons.

We have solved the BS equation by applying periodic boundary conditions that
restrict $\vec{k}$ to a regular discrete grid in the primitive zone illustrated in Fig.~\ref{band}.
Our main results are summarized in Fig.~\ref{Exciton}(a),
which shows the energies of Type III excitons as a function of center-of-mass momentum $\vec{Q}$.
To test the convergence of our calculations with respect to the $\vec{k}$-space
sampling density, we plot the binding energies of low-energy excitons
as a function of periodic system size in Fig.~\ref{Exciton}(b).
We start by analyzing excitons at and close to the $\Gamma$ point ($\vec{Q}=0$),
and then discuss the nearly degenerate low-energy excitons $\vec{Q}=K,K'$.

MoS$_2$ has a 3-fold rotational symmetry which can be used to classify
excitons with $\vec{Q}=0$: $\hat{C}_3|\chi\rangle=\exp(-i \frac{2\pi}{3}\mathcal{L} )|\chi\rangle$, where
the quantum number $\mathcal{L}$ takes on the discrete values $\mathcal{L} = -1, 0,1$.
An exciton state $|\chi_+\rangle$ with $\mathcal{L}=1$ has a time reversal (TR) partner $|\chi_-\rangle$ with the opposite $\mathcal{L}=-1$.
The combination of $\hat{C}_3$ and TR symmetries guarantees that
the TR pair $|\chi_+\rangle$ and $|\chi_-\rangle$ are degenerate in energy.
Breaking either $\hat{C}_3$ or TR symmetry can lift this degeneracy\cite{Dirac_Exciton, Cornell14, MIT14, Berkley14, Sriva14, Aiva14}.
The optical selection rule for circularly polarized light is related to $\hat{C}_3$ symmetry\cite{VP_Peking, MnPS3}:
$\langle \chi_+|\hat{j}_{-}|G\rangle=\langle \chi_-|\hat{j}_{+}|G\rangle=0$,
where $\hat{j}_{\pm}=\hat{j}_{x}\pm i\hat{j}_{y}$ is the current operator.
It follows that polarization-dependent optical studies can can be used to infer the $\mathcal{L}$
quantum number of a bright exciton.
$\mathcal{L}=0$ excitons are optically inactive $\langle \chi_0|\hat{j}_{\pm}|G\rangle=0$.

\begin{figure}[t]
 \includegraphics[width=1\columnwidth]{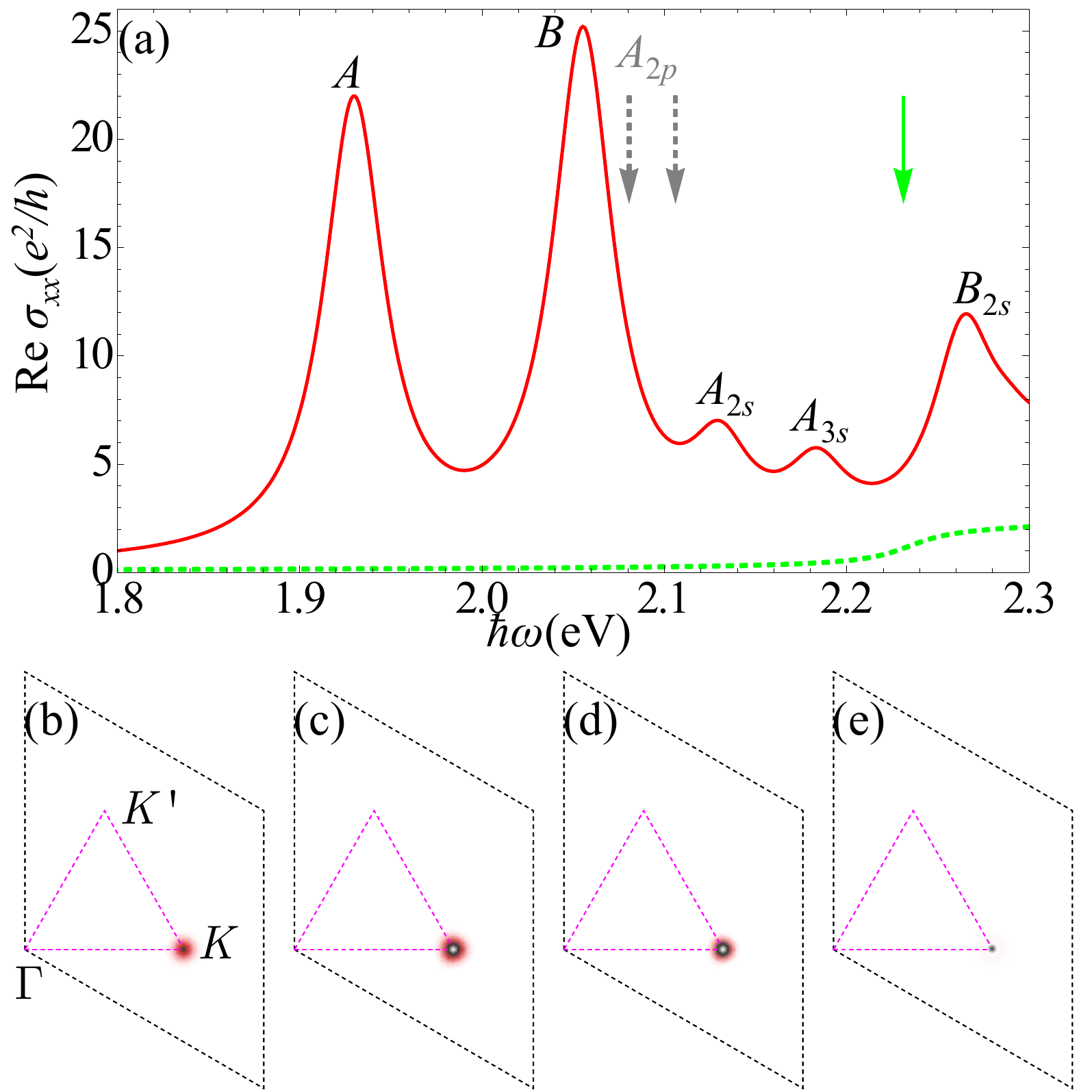}
 \caption{(Color online)
 (a) Real part of the optical conductivity with (solid red curve) and without (dashed green curve) electron-hole interactions.
 For these calculations the BS equation was solved on a $51\times51$ $\vec{k}$-grid and transitions were
 broadened by 20meV.   The solid green arrow indicates the quasiparticle band gap.
 The two dashed gray arrows mark the energies of the $A_{2p}$ excitons.
 Note that we have rigidly shift the excitation energy spectrum by a constant,
 so that the $A$ exciton energy is at 1.93eV as measured by photoluminescence experiments\cite{VP_Columbia, PL_Shih, PL_Rana, Trion_Columbia}.
 (b)-(e) $\vec{k}$-space maps of $P_{\vec{Q}=0}(\vec{k})$ for low-energy excitons.
 (b)The $\mathcal{L}=1$ exciton $|A_+\rangle$.
 (c)-(d) Two non-degenerate $A_{2p}$ excitons in valley $K$.
 (e) The $\mathcal{L}=1$ $A_{2s}$ exciton.
 }
 \label{cond}
\end{figure}

\subsection{Optical Conductivity}

Exciton states at $\vec{Q}=0$ are most easily studied experimentally because they
contribute to the optical conductivity.  Fig.~\ref{cond}(a) plots the real part of the in-plane optical conductivity which
has a number of clear features.
Peak $A$ stems from the doubly-degenerate excitons
expected from the symmetry analysis given above.
Fig.~\ref{cond}(b) illustrates $P_{\vec{Q}=0}(\vec{k})$ (see Eq.~(\ref{Prob})) for $|A_+\rangle$, the $\mathcal{L}=1$ exciton of peak $A$.
$|A_+\rangle$ is dominated by electron-hole transitions from valence band $v_1$ to conduction band
$c_2$ in valley $K$ (see Fig.~\ref{band}),
while its TR partner $|A_-\rangle$ is primarily composed of similar transitions in the opposite valley $K'$.
Excitons in the $B$ series are similar to those in $A$, and are dominated by
transitions from band $v_2$ to $c_1$ in valley $K$ ($K'$).
The lowest energy $A$ and $B$ excitons are analogous to the $1s$ states of a 2D hydrogenic model.
Fig.~\ref{cond}(a) also shows $2s$ and $3s$ peaks identified
in the $A$ series, and a $2s$ peak identified in the $B$ series.

Our calculation predicts that the lowest energy $A$ excitons have a binding energy $\sim 0.3$ eV,
in agreement with the estimate in Ref.~[\onlinecite{PL_Rana}].
As shown in Fig.~\ref{Exciton}(b), a $33\times33$ $\vec{k}$-grid already provides good
convergence for the lowest energy $A$ and $B$ excitons whereas, because
more weakly bound excitons have sharper structure
in momentum space as illustrated in Fig.~\ref{cond}(e), the $A_{2s}$ exciton requires a finer $\vec{k}$-grid for convergence.

The energies of $s$-wave excitons in the $A$($B$) series deviate strongly from the $(n-1/2)^{-2}$ pattern of 2D hydrogenic models.
This is partly due to the effective electron-hole interaction potential(Eq.~(\ref{potential})),
which differs from the standard Coulomb interaction because of the finite width of the TMD layer.
We also find that $A_{2p}$ states have a lower energy than $A_{2s}$ states.
This anomalous energy ordering is consistent with recent experimental and theoretical studies of monolayer WS$_2$\cite{Dark_WS2}.
More interestingly, the $A_{2p}$ states do \emph{not} have the 4-fold degeneracy expected in two-valley
systems. In fact, there are two non-degenerate $A_{2p}$ states within each valley, as illustrated in Fig.~\ref{cond}(c) and \ref{cond}(d).
This feature results from the dependence of band state wave functions on
momentum direction near $K$ and $K'$ valleys and is closely related to
similar properties of excitons in massive Dirac equation band models,
which we explain in detail in Appendix~\ref{AppB}.
The $A_{2p}$ states are not optically bright in one-photon spectra, but can be detected using two-photon techniques
like those achieved in recent experiments.\cite{Dark_WS2, Dark_WSe2}
We therefore expect that the energy splitting within the $A_{2p}$ states can be
experimentally measured.

\subsection{ Valley Coherence}

Valley coherence can be externally generated using linearly polarized light\cite{ValleyC},
and can also be intrinsically induced by electron-hole exchange interactions.
By treating finite $\vec{Q}$ terms in the BS equation(Eq.~(\ref{BS_matrix}))
as a first-order perturbation acting on valley-degenerate excitons,
we arrive at the effective Hamiltonian
\begin{equation}
\begin{aligned}
H_{\vec{Q}}^{\text{eff}}=&\big(\hbar \omega_0+ \frac{\hbar^2Q^2}{2M}\big) ] \, \tau_0+J_{\vec{Q}} \, \tau_0\\
                         +&J_{\vec{Q}}[\cos(2\phi_{\vec{Q}}) \tau_x + \sin(2\phi_{\vec{Q}}) \tau_y].
\end{aligned}
\label{H_eff}
\end{equation}
Here $\omega_0$ is the exciton energy at $\vec{Q}=0$, and
$M$ is the exciton mass. $\tau_0$ and $\tau_{x, y}$
are respectively identity and off-diagonal Pauli matrices in valley space.
$J_{\vec{Q}} \, \tau_0$ originates from intra-valley exchange interactions,
while inter-valley exchange interactions act as an in-plane pseudo-magnetic field in
the valley space and are captured by
the second line of Eq.~(\ref{H_eff}).
The dependence of $H_{\vec{Q}}^{\text{eff}}$ on $\phi_{\vec{Q}}$, the orientation angle of the 2D vector $\vec{Q}$,
follows from the wave-vector dependence of the conduction and valence
band states near $K$ and $K'$.
Ref.~[\onlinecite{Wu14}] and [\onlinecite{Dirac_Exciton}] have studied the inter-valley exchange interaction
and show that,
\begin{equation}
J_{\vec{Q}}\propto|\psi_{eh}(0)|^2 Q^2 V_{\vec{Q}},
\end{equation}
where $|\psi_{eh}(0)|^2$ is the probability that an electron and a hole overlap spatially.
In the small $Q$ limit, $V_{\vec{Q}}\propto 1/(Q(1+r_0 Q))$ for the potential in Eq.~(\ref{potential}).
Therefore, $J_{\vec{Q}}$ scales linearly with $Q$ in the long wave-length limit.
We note that while inter-valley exchange interaction endows finite-momentum excitons with chirality $I=2$
as pointed out by Ref.~[\onlinecite{Dirac_Exciton}], intra-valley exchange\cite{Wu14, Aiva14, Yuri15}
is also important especially in regard to the exciton energy dispersion.

Equation~(\ref{H_eff}) is derived from the massive Dirac model approximation to the quasiparticle band structure near K and K'.\cite{Wu14, Dirac_Exciton}
Our lattice calculation verifies this low-energy effective theory.
Fig.~\ref{Exciton}(a) shows that there are two non-degenerate energy branches which evolve from
the double-degenerate $\vec{Q}=0$ $A$, $B$ and $A_{2s}$ excitons.
In each energy branch, the exciton state is a coherent superposition of
direct excitons at the two valleys, as demonstrated in the left inset of Fig.~\ref{Exciton}(a).
The lower and upper energy branch have respectively quadratic and linear dispersion
in the long wave-length limit, in agreement with the prediction of Eq.~(\ref{H_eff}).
Unlike their $s$-wave cousins, branches evolving from
valley-degenerate $\vec{Q}=0$ $A_{2p}$ excitons
remain doubly degenerate at finite momentum
because $|\psi_{eh}(0)|^2$ is zero for $p$-wave excitons and exchange interactions
therefore vanish.

Photons with the energy of an $A$ or $B$ exciton
can at most provide a momentum with magnitude $\sim \omega_{A (B)}/c$, where $c$ is the speed of light.
According to our calculation, this momentum corresponds to an energy splitting of
$0.4 (0.5)$ meV between the two energy branches evolving from $A$($B$),
and a period of 11(8)ps for Rabi oscillation between the two valleys.
We conclude that interaction-induced valley coherence
provides an important mechanism for valley depolarization\cite{Urbaszek14, Wu14, Marie14},
in addition to that provided by impurity scattering.

\subsection{Brillouin-zone corner Excitions}

Large-momentum excitons composed of electrons and holes in opposite valleys
can have an energy similar to those with zero momentum.
In Fig.~\ref{Exciton}(a), the $\chi_K^2$ and $\chi_K^4$ excitons
have center-of-mass momentum $\vec{Q}= K$ when reduced to the first Brillouin-zone.
They are dominated respectively by transitions between valence band holes of $v_1$ and
$v_2$ states in valley $K$  and conduction band electrons of  $c_1$ and $c_2$ states in the opposite valley
$K'$.  The $\chi_K^2$ and $\chi_K^4$ excitons are the $\vec{Q}=K$
counterparts of the $A$ and $B$ excitons and have nearly the same energy.
The small differences have two origins, the energy splitting between $c_1$ and $c_2$ bands\cite{c1c2},
and a change in the exchange interaction.

\begin{figure}[t]
 \includegraphics[width=0.9\columnwidth]{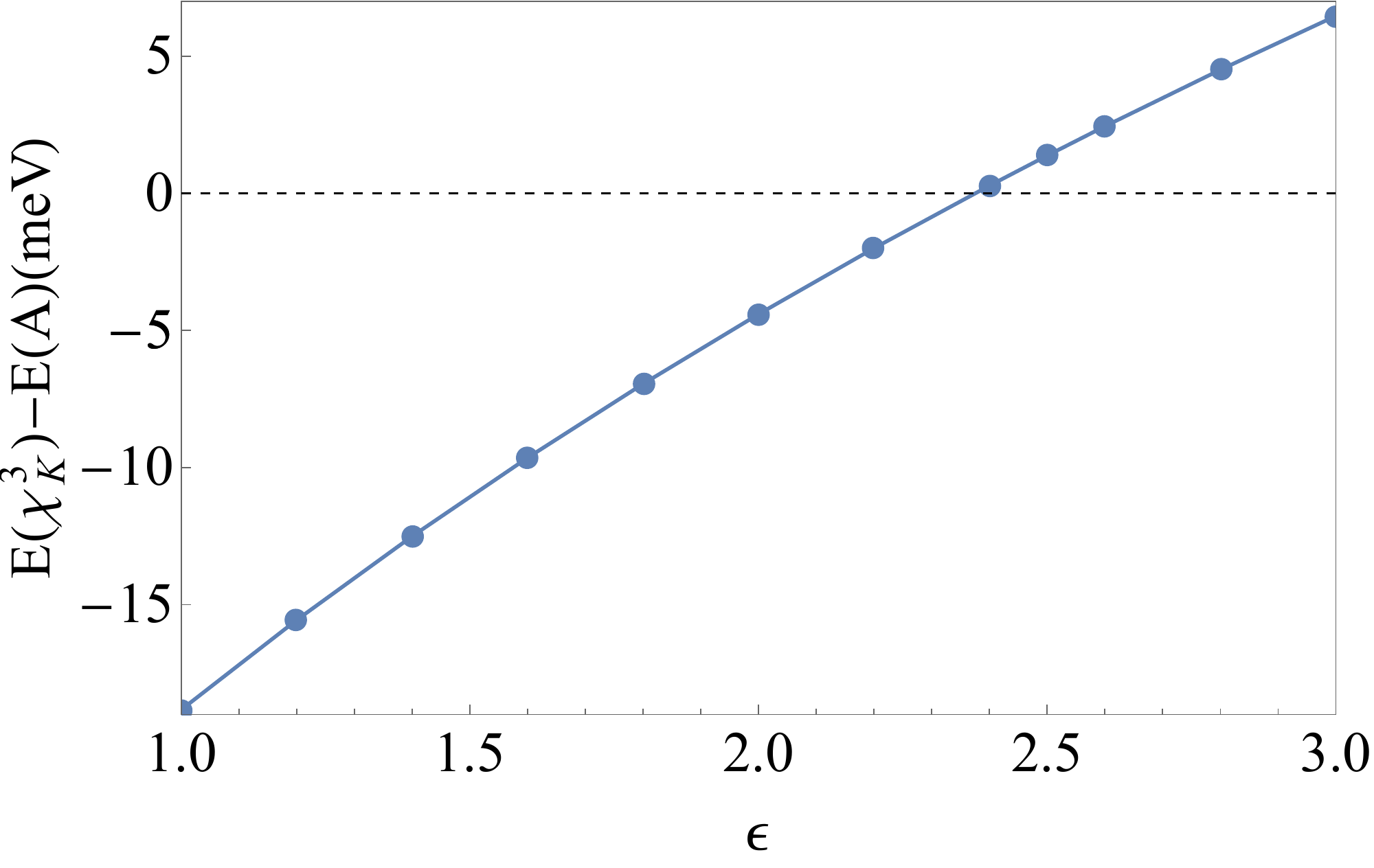}
 \caption{The energy difference between the large-momentum exciton $\chi_K^3$ and zero-momentum exciton $A$ as a function of dielectric constant $\epsilon$.
  The calculation was performed on a $45\times 45$ $\vec{k}$-grid. $\chi_K^3$ is the triplet exciton state with a hole in $\Gamma$ valley and an electron in $K$ valley.}
 \label{splitting}
\end{figure}

However another set of excitations appear at the same crystal momentum.
Low energy excitations at $\vec{Q}=K$ also originate from
holes in the $\Gamma$ valley and electrons in the $K$ valley, as illustrated in the right inset of Fig.~\ref{Exciton}(a).
In our lattice calculations $\chi_K^1$ and $\chi_K^3$ are such excitons.
If we neglect the spin splitting of both topmost valence bands at $\Gamma$
and the lowest conduction bands at the $K$ valley, the corresponding excitons can be
classified as singlets and triplets according to their spin configurations.
Singlets experience strong electron-hole exchange interactions, because the
valence band maximum at $\Gamma$ and the
conduction band minimum at $K$ have the same $d_{z^2}$ orbital character.
The exchange energy is proportional to $V_{\vec{Q}=K}=(U-1.28)V_{R=a_0}$.
The exchange interactions effectively vanishes for triplets because of their particular spin structure.
In our calculation, $U$ is set to be $1$ and the exchange energy for singlets is therefore negative.
Therefore, we can identify $\chi_K^1$ as the singlet state and $\chi_K^3$ as one of the triplet states.
Depending on the value of $U$ used in our model, the lowest energy excitons can
occur at $\vec{Q}=0$, corresponding to a direct gap system, or at $\vec{Q}=K,K'$
corresponding to an indirect gap system.  Because the appropriate value which should be used for
$U$ depends on electronic correlations at the atomic level and on screening by remote bands
not included in our calculation, we are not able to reach a definitive conclusion as to
whether or not the singlet $\chi_K^1$ is the lowest energy exciton.
However, the energy of the triplet $\chi_K^3$ does not suffer from such uncertainty.
The valence band effective mass at the $\Gamma$ point is heavier than that at the $K$ point\cite{Gamma}.
Electron-hole pairs are therefore bound more strongly in $\chi_K^3$ than in the $A$ exciton,
which compensates for the energy difference (68meV) between the topmost valence bands at $\Gamma$ and $K$ points,
and makes the energies of the triplet $\chi_K^3$ and $A$ excitons very close to each other.
By decreasing the dielectric constant $\epsilon$ to 1, we find that the triplet $\chi_K^3$ becomes lower in energy than
the $A$ exciton by 19meV, as illustrated in Fig.~\ref{splitting}.  The energetic ordering of direct and indirect exciton states in
single-layer TMDs could therefore depend on the two-dimensional system's three-dimensional
dielectric environment.

\section{Summary and Conclusions}

In summary, we have constructed a lattice model based on Mo $d$-orbitals to study exciton band sturcture of monolayer MoS$_2$.
Zero-momentum excitons have non-hydrogenic energy series, because screening effect has a spatial dependence,
and band edges in $K$ and $K'$ valley are described by massive Dirac equation.
The energy splitting within $A_{2p}$ excitons remains to be measured using methods such as two-photon technique.
The exciton band structure exhibits non-degenerate energy branches evolving from valley-degenerate bright excitons, indicating valley coherence.
Such low-energy branches are well separated from the continuum spectrum, which justifies the application of low-energy effective theory\cite{Wu14, Dirac_Exciton}. We find that low-energy excitons can possess a large momentum, either with electron and hole in opposite valleys ($K$, $K'$), or with hole in $\Gamma$ valley and electron in $K$($K'$) valley.
Large-momentum low-energy exciton states can provide relaxation channels for bright excitons, and reduce photoluminescence quantum efficiency.

\section{Acknowledgments}
This work was supported by the DOE Division of Materials Sciences
and Engineering under grant DE-FG03-02ER45958, and
by the Welch foundation under grant TBF1473.
We thank Texas Advanced Computing Center(TACC)
for providing computer time allocations.

\appendix
\section{Explicit form of tight-binding model}
\label{AppA}

We approximate the quasiparticle-Hamiltonian matrix $\mathcal{H}_{\vec{k}}$
by a tight-binding model which generalizes Ref.~[\onlinecite{ThreeB}] from three to five
d-bands:
\begin{equation}
\mathcal{H}_{\vec{k}}=\lambda\vec{L}\cdot\vec{S}+I_2\otimes\mathcal{H}_0(\vec{k}).
\label{tightB}
\end{equation}
The first term $\lambda\vec{L}\cdot\vec{S}$ describes the on-site atomic spin-orbit coupling of Mo $d$ orbitals,
where $\vec{L}$ and $\vec{S}$ are respectively the orbital and spin angular momentum.
The coupling constant $\lambda=0.073$eV, which was adjusted to fit the valence-band spin splitting at the $K$ point as detailed in Ref.~[\onlinecite{ThreeB}].
The second term $I_2\otimes\mathcal{H}_0(\vec{k})$ is spin independent,
where $I_2$ is a $2\times2$ identity matrix in spin space,
and $\mathcal{H}_0(\vec{k})$ is a $5\times5$ matrix in orbital space.
Because of the mirror symmetry with respect to the Mo plane,
$\mathcal{H}_0(\vec{k})$ is block-diagonal:
\begin{equation}
\mathcal{H}_0(\vec{k})=\begin{pmatrix}
  \mathcal{H}_0^{\text{even}}(\vec{k}) & 0 \\
  0 & \mathcal{H}_0^{\text{odd}}(\vec{k})
 \end{pmatrix}.
\end{equation}
$\mathcal{H}_0^{\text{even}}(\vec{k})$ is a $3\times3$ matrix in the bases
$\{|d_{z^2}\rangle, |d_{xy}\rangle, |d_{x^2-y^2}\rangle\}$.
Similarly, $\mathcal{H}_0^{\text{odd}}(\vec{k})$ is a $2\times2$ matrix in the bases
$\{|d_{xz}\rangle, |d_{yz}\rangle\}$.

For $\mathcal{H}_0^{\text{even}}(\vec{k})$, we adopt a model constructed in Ref.~\onlinecite{ThreeB}.
The construction uses point-group symmetries to minimize the number of parameters,
and the parameters are fitted from first-principle energy bands.
An explicit form of $\mathcal{H}_0^{\text{even}}(\vec{k})$ with hoppings up to third-nearest-neighbor
is given in Eqs.~(13) to (24) of Ref.~\onlinecite{ThreeB}.
We generalize the symmetry-based method used in Ref.~\onlinecite{ThreeB} to
construct the remaining part of the Hamiltonian $\mathcal{H}_0^{\text{odd}}(\vec{k})$, and its form is:
\begin{equation}
\mathcal{H}_0^{\text{odd}}(\vec{k})=\begin{pmatrix}
  h_{x}(\vec{k}) & h_{xy}(\vec{k}) \\
  h_{xy}^{*}(\vec{k}) & h_{y}(\vec{k})
 \end{pmatrix},
\end{equation}
in which
\begin{equation}
\begin{aligned}
h_{x}(\vec{k})=&O_1+2t\cos2\alpha+(t+3t')\cos\alpha\cos\beta\\
+&4s\cos3\alpha\cos\beta+(3s'-s)\cos2\beta\\
+&2u\cos4\alpha+(u+3u')\cos2\alpha\cos2\beta,\\
h_{y}(\vec{k})=&O_1+2t'\cos2\alpha+(t'+3t)\cos\alpha\cos\beta\\
+&4s'\cos3\alpha\cos\beta+(3s-s')\cos2\beta\\
+&2u'\cos4\alpha+(u'+3u)\cos2\alpha\cos2\beta,\\
h_{xy}(\vec{k})=& 4it_{xy}\sin\alpha(\cos\alpha-\cos\beta)\\
+&\sqrt{3}(t'-t)\sin\alpha\sin\beta\\
+&2\sqrt{3}(s'-s)\sin\alpha\sin\beta(1+2\cos2\alpha)\\
+&4iu_{xy}\sin2\alpha(\cos2\alpha-\cos2\beta)\\
+&\sqrt{3}(u'-u)\sin2\alpha\sin2\beta,\\
(\alpha,\beta)=&(\frac{1}{2}k_xa_0,\frac{\sqrt{3}}{2}k_ya_0).
\end{aligned}
\end{equation}
Hoppings in real space up to third nearest neighbors are included.
The numerical value of the parameters in unit of eV is:
\begin{equation}
\begin{aligned}
O_1=&3.558,\\
t=&-0.189, t'=-0.117, t_{xy}=0.024,\\
s=&-0.041, s'=0.003,\\
u=&0.165, u'=-0.122, u_{xy}=-0.140,
\end{aligned}
\end{equation}
which are obtained by fitting to first-principle calculations as shown in Fig.~\ref{band}.

$\mathcal{H}_{\vec{k}}$ is diagonalized by the unitary matrix $\mathcal{U}_{\vec{k}}$ (Eq.~(\ref{DX})),
and the corresponding eigenvalues are quasiparticle energies $\varepsilon_{\vec{k}n}$.

\section{Massive Dirac model for excitons}
\label{AppB}
In this appendix we study excitons in the massive Dirac model,
and show that this simple model captures many important features of the $\vec{Q}=0$ excitons in monolayer MoS$_2$.
In the vicinity of $K$ or $K'$ point, the $k\cdot p$ Hamiltonian for two bands ($c_2$, $v_1$)
is described by the massive Dirac model\cite{Yao12}:
\begin{equation}
\mathcal{H}_\tau(\vec{k})=\hbar v_F k[\cos(\Phi_{\vec{k}})\sigma_x+\sin(\Phi_{\vec{k}})\sigma_y]+\Delta \sigma_z,
\label{Htau}
\end{equation}
where $\sigma_{x, y, z}$ is the Pauli matrices for the basis function at $K$ or $K'$ point, and $\Delta$ is the energy gap.
The angle $\Phi_{\vec{k}}$ is defined as $\cot{\Phi_{\vec{k}}}=\tau k_{x}/{k_y}$, where $\tau=\pm1$ labels valley $K$ and $K'$.
The conduction and valence band are described by spinors:
\begin{equation}
|c, \vec{k}\rangle_\tau=\begin{pmatrix} \cos(\frac{\theta_{k}}{2})\\\sin(\frac{\theta_{k}}{2})e^{i\Phi_{\vec{k}}}\end{pmatrix},
|v, \vec{k}\rangle_\tau=\begin{pmatrix} \sin(\frac{\theta_{k}}{2})e^{-i\Phi_{\vec{k}}} \\ -\cos(\frac{\theta_{k}}{2})\end{pmatrix},
\end{equation}
where angle $\theta_{k}$ is defined as $\cos{\theta_{k}}=\Delta/\varepsilon_{k}$ with $\varepsilon_{k}=\sqrt{\Delta^2+(\hbar v_F k)^{2}}$.

\begin{figure}[t]
 \includegraphics[width=0.85\columnwidth]{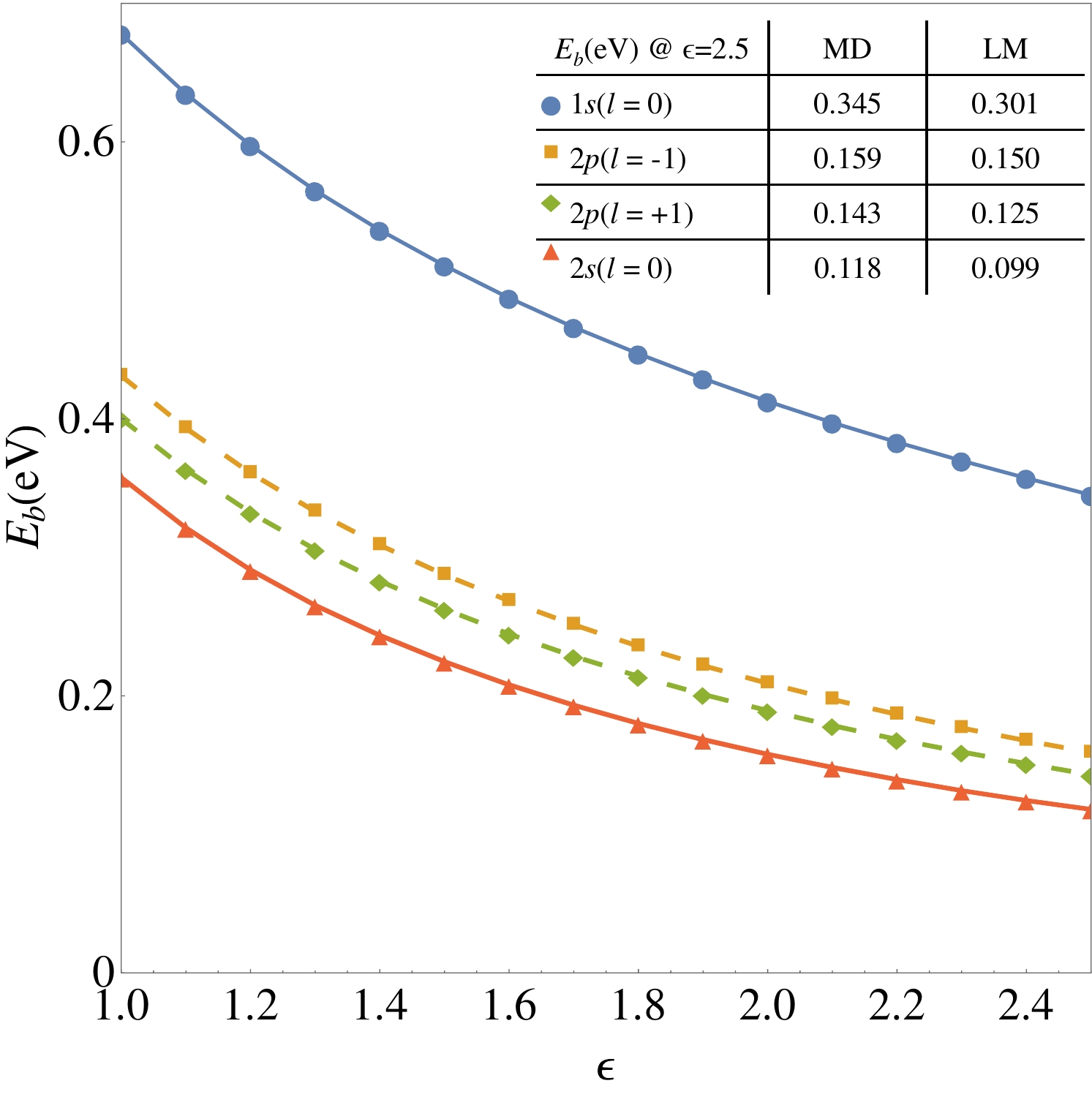}
 \caption{(Color online) Binding energy $E_b$ in massive Dirac model
  for $1s$, $2s$, and $2p$ excitons in valley $K$ ($\tau=1$) as a function of dielectric constant $\epsilon$.
  The parameter values are
$\hbar v_F=1.105\text{eV}\times3.193\text{\AA}$, $\Delta=0.7925\text{eV}$ and $r_0=33.875\text{\AA}/\epsilon$.
 The inset table compares $E_b$ obtained respectively from massive Dirac model (MD) and lattice model (LM) for $\epsilon=2.5$.}
 \label{MDirac}
\end{figure}

As discussed in the main text, inter-valley coupling is nearly absent for $\vec{Q}=0$ excitons.
Within each valley, the kernel of the BS equation in Eq.~(\ref{BS_matrix}) can be expressed in terms of band spinors:
\begin{equation}
\begin{aligned}
\mathcal{K}_\tau(\vec{k}, \vec{k}')=&\delta_{\vec{k} \vec{k}'}T_{k}-D_\tau(\vec{k},\vec{k}'),\\
T_{k}=&2\varepsilon_{k},\\
D_\tau(\vec{k},\vec{k}')=&\frac{1}{\mathcal{A}}\tilde{V}_{\vec{k}-\vec{k}'} \Big({}_\tau\langle c, \vec{k}|c, \vec{k}'\rangle_\tau
                                                                  {}_\tau\langle v, \vec{k}'|v, \vec{k}\rangle_\tau \Big)\\
                   =&\frac{1}{4\mathcal{A}}\tilde{V}_{\vec{k}-\vec{k}'}\big[(1+\cos{\theta_{k}})(1+\cos{\theta_{k'}})\\
                   &+2\sin{\theta_{k}}\sin{\theta_{k'}}e^{i\tau (\phi_{\vec{k}'}-\phi_{\vec{k}})}\\
                   &+(1-\cos{\theta_{k}})(1-\cos{\theta_{k'}})e^{i2\tau (\phi_{\vec{k}'}-\phi_{\vec{k}})}\big],
\end{aligned}
\label{KernelT}
\end{equation}
where $T_{k}$ can be understood as the kinetic energy.
$D_\tau(\vec{k},\vec{k}')$ is the electron-hole direct interaction,
while the exchange interaction $X$ (Eq.~(\ref{BS_matrix})) is neglected.
$\mathcal{A}$ is the area of the $2D$ system,
and $\phi_{\vec{k}}$ is the orientation angle of $\vec{k}$ with $\cot{\phi_{\vec{k}}}=k_{x}/{k_y}$.
In Eq.~(\ref{KernelT}), angle $\phi_{\vec{k}}$ is used
instead of $\Phi_{\vec{k}}$ (Eq.~(\ref{Htau})) so that the valley dependence of $\mathcal{K}_\tau$ is explicit.
The interaction potential has the following form
\begin{equation}
\tilde{V}_{q}=\frac{2\pi e^2}{\epsilon q} F(q),
\label{Fq}
\end{equation}
where the form factor $F(q)=1/(1+r_0 q)$ modifies the Coulomb interaction in consistency with the real-space interaction potential (Eq.~(\ref{potential})).
The BS equation reads:
\begin{equation}
\sum_{\vec{k}'}\mathcal{K}_{\tau}(\vec{k}, \vec{k}')\psi(\vec{k}')=E\psi(\vec{k}).
\end{equation}

\begin{figure}[t]
 \includegraphics[width=0.85\columnwidth]{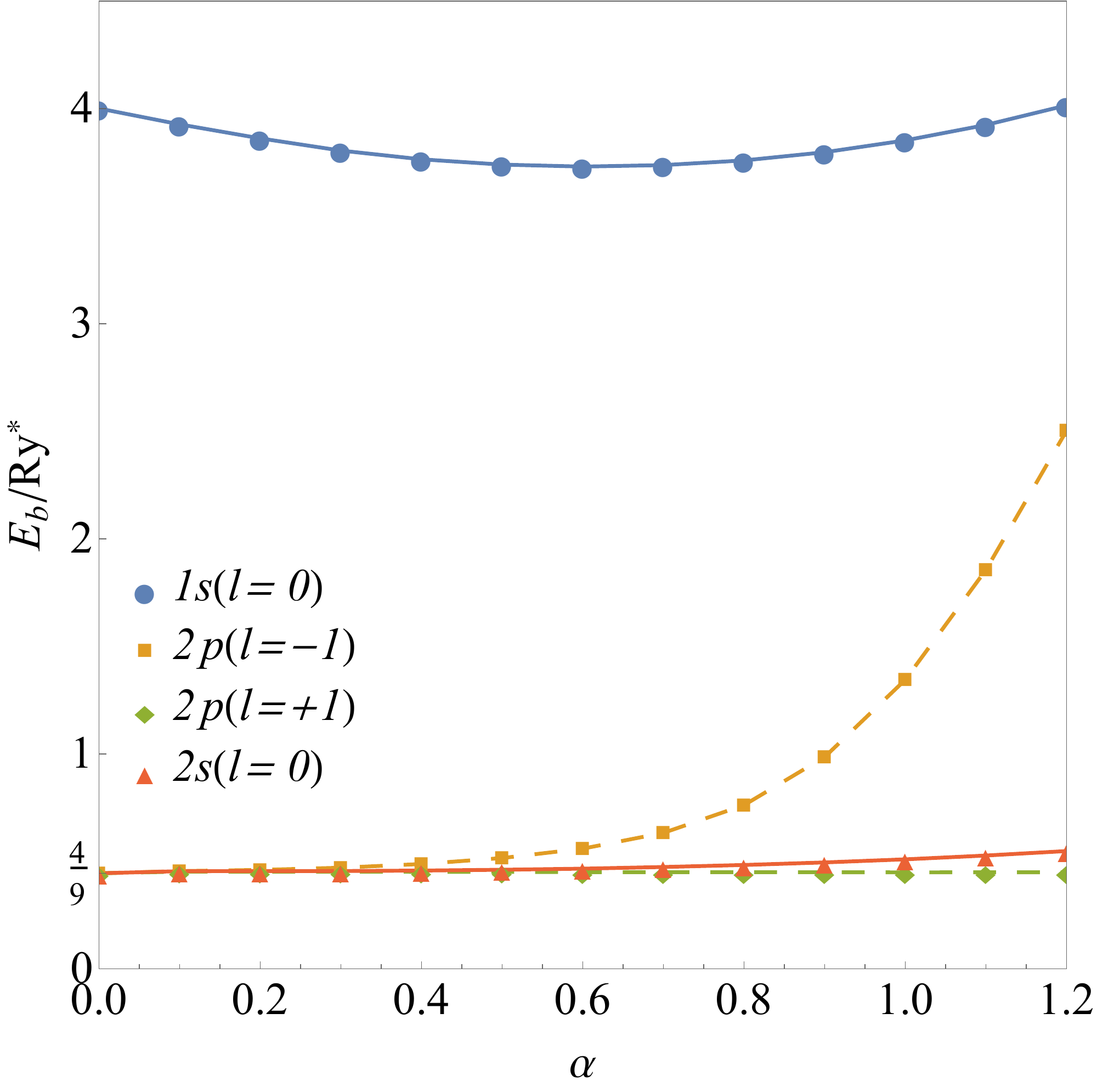}
 \caption{(Color online) Binding energy $E_b$ in massive Dirac model with standard Coulomb interaction($r_0=0, F(q)=1$)
  for $1s$, $2s$, and $2p$ excitons in valley $K$ ($\tau=1$) as a function of fine structure constant $\alpha$.
  %$E_b$ is in unit of effective Rydberg energy $Ry^*$ in this figure.
  }
 \label{Dirac_Coulomb}
\end{figure}

We can define effective Bohr radius $a_B^*$, Rydberg energy $Ry^*$ and fine structure constant $\alpha$:
\begin{equation}
\begin{aligned}
a_B^*&=\frac{2 \epsilon (\hbar v_F)^2}{e^2 \Delta},
Ry^*=\frac{1}{2}\frac{e^2}{\epsilon a_B^*},
\alpha=\frac{e^2}{\epsilon \hbar v_F} .
\end{aligned}
\end{equation}
For notation convenience, we also define parameter $\beta=(\alpha/2)^2$.

After taking $a_B^*$ as unit of length and $Ry^*$ as unit of energy,  and using an ansatz $\psi(\vec{k})=\psi(k)e^{i l \phi_{\vec{k}}}$,
BS equation is reduced to the following 1D eigenvalue problem:
\begin{equation}
\begin{aligned}
&E\psi(k)=T_{k}\psi(k)\\
-&\int_0^{\infty}dk'
\big[(1+\cos{\theta_{k}})(1+\cos{\theta_{k'}}) I_{kk'}(l)\\
+&2\sin{\theta_{k}}\sin{\theta_{k'}} I_{kk'}(l+\tau )\\
+&(1-\cos{\theta_{k}})(1-\cos{\theta_{k'}}) I_{kk'}(l+2\tau )
\big]k'\psi(k'),
\end{aligned}
\label{1D}
\end{equation}
where $T_{k}$, $\cos{\theta_{k}}$ and $I_{kk'}(l)$ in effective atomic units are:
\begin{equation}
\begin{aligned}
T_k&=\frac{2}{\beta}\sqrt{1+\beta k^{2}}, \cos\theta_{k}=1/\sqrt{1+\beta k^{2}},\\
I_{kk'}(l)&=\frac{1}{4\pi}\int_{0}^{2\pi}d\phi \frac{F(\sqrt{k^2+k'^2-2kk'\cos{\phi}})\cos{l\phi}}{\sqrt{k^2+k'^2-2kk'\cos{\phi}}}.
\end{aligned}
\label{Expr}
\end{equation}

Eq.~(\ref{1D}) makes it clear that within the same valley excitons with quantum number $l$ and $-l$ are not degenerate in energy,
because of the wave-vector dependence of the band spinors.
However, there is a degeneracy between ($\tau$, $l$) and (-$\tau$, -$l$) excitons, which originates from time reversal symmetry.

We apply the massive Dirac model to excitons in $A$ series of monolayer MoS$_2$. The appropriate parameter values are
$\hbar v_F=1.105\text{eV}\times3.193\text{\AA}$, $\Delta=0.7925\text{eV}$\cite{ThreeB} and $r_0=33.875\text{\AA}/\epsilon$\cite{PL_Rana}.
The effective atomic units then take the following value,
$a_B^*=\epsilon\times 2.18\text{\AA}$ and $Ry^*=3.3\text{eV}/\epsilon^2$,
and the fine structure constant $\alpha=4.075/\epsilon$.
Eq.~(\ref{1D}) is solved numerically by discretizing the 1D $k$-space.
The result is presented in Fig.~\ref{MDirac},
which depicts the binding energy of $1s$, $2s$, and $2p$ excitons in valley $K$ as a function of dielectric constant $\epsilon$.
It reproduces all essential features of $\vec{Q}=0$ excitons discussed in the main text:
(1)binding energies deviate from the pattern of the 2D hydrogen model;
(2)$2p$ states have a larger binding energy than $2s$;
and (3)there is an energy splitting between $l=\pm1$ $2p$ states within the same valley.
Moreover, the binding energies calculated by using the Massive Dirac model and the lattice model of the main text are close to each other as shown in the inset table of Fig.~\ref{MDirac}.

Finally, we study the massive Dirac model with standard Coulomb interaction by taking form factor $F(q)$ to be 1.
In this case, the fine structure constant $\alpha$ controls the deviation of the massive Dirac model from the 2D hydrogen model.
Fig.~\ref{Dirac_Coulomb} presents the binding energy as a function of $\alpha$.
In the limit of weak electron-hole interaction ($\alpha \rightarrow 0$), the massive Dirac model reduces to the 2D hydrogen model
as implied by Eq.~(\ref{1D}) and (\ref{Expr}).
Therefore, $E_b(1s)=4Ry^*$ and $E_b(2s)=E_b(2p)=\frac{4}{9}Ry^*$ as $\alpha$ goes to 0.
$2s$ and $2p$ states remain nearly degenerate for $\alpha<0.4$,
and develop prominent energy splitting at large $\alpha$.
The energy ordering is $E_b(2p, l= -1)> E_b(2s)>E_b(2p, l= +1)$ in valley $K$.
Note in the case of monolayer MoS$_2$ where interaction potential is modified by $F(q)$ (Eq.~(\ref{Fq})),
all $2p$ states have bigger binding energies than $2s$ states as shown in Fig.~\ref{MDirac}.


\begin{thebibliography}{}
\bibitem{mirror}
Under mirror operation $\mathcal{M}$,
a vector is transformed as $\mathcal{M}(x, y, z)=(x, y, -z)$;
for a spinor in spin space $\mathcal{M}|s\rangle=\exp[-i\pi s_z/2]|s\rangle$
with $s_z$ being $z$ component Pauli matrix in spin space.
\end{thebibliography}
\end{document}